\documentclass{aastex}
\usepackage{amsmath}
\usepackage{graphicx}
\usepackage{url}
\usepackage{amssymb}
\usepackage{float}
\makeatletter
\let\@dates\relax
\makeatother

\usepackage[ps2pdf,%
a4paper=true,%
breaklinks=true,%
colorlinks=true,%
citecolor=blue,
pdfauthor={Bruno et al.},%
pdftitle={Geomagnetically trapped, albedo and solar energetic particles: trajectory analysis and flux reconstruction with PAMELA}%
]{hyperref}

\begin{document}

\title{Geomagnetically trapped, albedo and solar energetic particles: trajectory analysis and flux reconstruction with PAMELA}
\author{
A.~Bruno$^{1,2,*}$,
O.~Adriani$^{3,4}$,
G.~C.~Barbarino$^{5,6}$,
G.~A.~Bazilevskaya$^{7}$,
R.~Bellotti$^{1,2}$,
M.~Boezio$^{8}$,
E.~A.~Bogomolov$^{9}$,
M.~Bongi$^{3,4}$,
V.~Bonvicini$^{8}$,
S.~Bottai$^{4}$,
F.~Cafagna$^{2}$,
D.~Campana$^{6}$,
P.~Carlson$^{11}$,
M.~Casolino$^{12,13}$,
G.~Castellini$^{14}$,
E.~C.~Christian$^{15}$,
C.~De~Donato$^{12,17}$,
G.~A.~de~Nolfo$^{15}$,
C.~De~Santis$^{12,17}$,
N.~De~Simone$^{12}$,
V.~Di~Felice$^{12,18}$,
A.~M.~Galper$^{16}$,
A.~V.~Karelin$^{16}$,
S.~V.~Koldashov$^{16}$,
S.~Koldobskiy$^{16}$,
S.~Y.~Krutkov$^{9}$,
A.~N.~Kvashnin$^{7}$,
A.~Leonov$^{16}$,
V.~Malakhov$^{16}$,
L.~Marcelli$^{12,17}$,
M.~Martucci$^{17,20}$,
A.~G.~Mayorov$^{16}$,
W.~Menn$^{21}$,
M.~Merg\`e$^{12,17}$,
V.~V.~Mikhailov$^{16}$,
E.~Mocchiutti$^{8}$,
A.~Monaco$^{1,2}$,
N.~Mori$^{3,4}$,
R.~Munini$^{8,19}$,
G.~Osteria$^{6}$,
F.~Palma$^{12,17}$,
B.~Panico$^{6}$,
P.~Papini$^{4}$,
M.~Pearce$^{11}$,
P.~Picozza$^{17}$,
M.~Ricci$^{20}$,
S.~B.~Ricciarini$^{4,14}$,
J.~M.~Ryan$^{10}$,
R.~Sarkar$^{22,23}$,
V.~Scotti$^{5,6}$,
M.~Simon$^{21}$,
R.~Sparvoli$^{12,17}$,
P.~Spillantini$^{16,24}$,
S.~Stochaj$^{25}$,
Y.~I.~Stozhkov$^{7}$,
A.~Vacchi$^{8}$,
E.~Vannuccini$^{4}$,
G.~I.~Vasilyev$^{9}$,
S.~A.~Voronov$^{16}$,
Y.~T.~Yurkin$^{16}$,
G.~Zampa$^{8}$
and N.~Zampa$^{8}$.
}

\affil{$^{1}$ Department of Physics, University of Bari, I-70126 Bari, Italy.}
\affil{$^{2}$ INFN, Sezione di Bari, I-70126 Bari, Italy.}
\affil{$^{3}$ Department of Physics and Astronomy, University of Florence, I-50019 Sesto Fiorentino, Florence, Italy.}
\affil{$^{4}$ INFN, Sezione di Florence, I-50019 Sesto Fiorentino, Florence, Italy.}
\affil{$^{5}$ Department of Physics, University of Naples ``Federico II'', I-80126 Naples, Italy.}
\affil{$^{6}$ INFN, Sezione di Naples, I-80126 Naples, Italy.}
\affil{$^{7}$ Lebedev Physical Institute, RU-119991 Moscow, Russia.}
\affil{$^{8}$ INFN, Sezione di Trieste, I-34149 Trieste, Italy.}
\affil{$^{9}$ Ioffe Physical Technical Institute, RU-194021 St. Petersburg, Russia.}
\affil{$^{10}$ Space Science Center, University of New Hampshire, Durham, NH, USA.}
\affil{$^{11}$ KTH, Department of Physics, and the Oskar Klein Centre for Cosmoparticle Physics, AlbaNova University Centre, SE-10691 Stockholm, Sweden.}
\affil{$^{12}$ INFN, Sezione di Rome ``Tor Vergata'', I-00133 Rome, Italy.}
\affil{$^{13}$ RIKEN, Advanced Science Institute, Wako-shi, Saitama, Japan.}
\affil{$^{14}$ IFAC, I-50019 Sesto Fiorentino, Florence, Italy.}
\affil{$^{15}$ Heliophysics Division, NASA Goddard Space Flight Ctr, Greenbelt, MD, USA.}
\affil{$^{16}$ National Research Nuclear University MEPhI, RU-115409 Moscow, Russia.}
\affil{$^{17}$ Department of Physics, University of Rome ``Tor Vergata'', I-00133 Rome, Italy.}
\affil{$^{18}$ Agenzia Spaziale Italiana (ASI) Science Data Center, I-00133 Rome, Italy.}
\affil{$^{19}$ Department of Physics, University of Trieste, I-34147 Trieste, Italy.}
\affil{$^{20}$ INFN, Laboratori Nazionali di Frascati, I-00044 Frascati, Italy.}
\affil{$^{21}$ Department of Physics, Universit\"{a}t Siegen, D-57068 Siegen, Germany.}
\affil{$^{22}$ Indian Centre for Space Physics, Kolkata 700084, West Bengal, India.}
\affil{$^{23}$ Previously at INFN, Sezione di Trieste, I-34149 Trieste, Italy.}
\affil{$^{24}$ IAPS/INAF, I-00133 Rome, Italy.}
\affil{$^{25}$ Electrical and Computer Engineering, New Mexico State University, Las Cruces, USA.}
\affil{$^{*}$ Corresponding author. E-mail address: alessandro.bruno@ba.infn.it.}

\begin{abstract}
The PAMELA satellite experiment is providing comprehensive observations of the interplanetary and ma\-gne\-to\-sphe\-ric radiation in the near-Earth environment. Thanks to its identification capabilities and the semi-polar orbit, PAMELA is able to precisely measure the energetic spectra and the angular distributions of the different cosmic-ray populations over a wide latitude region, including geo\-ma\-gne\-tically trapped and albedo particles. Its observations comprise the solar energetic particle events between solar cycles 23 and 24, and the geo\-ma\-gne\-tic cutoff variations during ma\-gne\-to\-sphe\-ric storms. PAMELA's measurements are supported by an accurate analysis of particle trajectories in the Earth's ma\-gne\-to\-sphe\-re based on a realistic geo\-ma\-gne\-tic field modeling, which allows the classification of particle populations of different origin and the investigation of the asymptotic directions of arrival.
\end{abstract}


\section{Introduction}
PAMELA (a Payload for Antimatter Matter Exploration and Light-nuclei Astrophysics) is a space-based experiment designed for a precise measurement of charged Cosmic-Rays (CR) -- protons, electrons, their antiparticles and light nuclei -- in the kinetic energy interval from several tens of MeV up to several hundreds of GeV \citep{PHYSICSREPORTS}. The Resurs-DK1 satellite, which hosts the apparatus, was launched into a semi-polar (70 deg inclination) and elliptical (350--610 km altitude) orbit on 15 June 2006; in 2010 it was changed to an approximately circular orbit at an altitude of $\sim$ 580 km. The instrument consists of a magnetic spectrometer equipped with a si\-li\-con tracking system, a time-of-flight system shielded by an anticoincidence system, an electromagnetic calorimeter and a neutron detector. Details about apparatus performance, proton selection, detector efficiencies and experimental uncertainties can be found elsewhere (see e.g. \citet{SOLARMOD}).

PAMELA is providing comprehensive observations of the interplanetary \citep{SOLARMOD,ELECTRONS} and ma\-gne\-to\-sphe\-ric \citep{PAMELAtrappedpbars,PAMTRAPPED,ALBEDO,GSTORM} radiation in the near-Earth environment. In particular, PA\-ME\-LA is able to precisely measure the Solar Energetic Particle (SEP) events du\-ring solar cycles 23 and 24 \citep{SEP2006,MAY17PAPER}. This work reviews PAMELA's main ma\-gne\-to\-sphe\-ric results, with the focus on the ana\-ly\-sis methods developed to support the observations, based on an accurate reconstruction of particle trajectories in the Earth's ma\-gne\-to\-sphe\-re.

%
%

\section{Geomagnetically Trapped and Re-Entrant Albedo Protons}
The Van Allen belts constitute a major radiation source in the Earth's vi\-ci\-ni\-ty. Speci\-fi\-cally, the inner belt is mainly populated by energetic protons, mostly originated by the decay of albedo neutrons according to the CRAND mechanism \citep{CRAND}, experiencing long-term magnetic trapping. Despite the significant improvement made in the latest decades, the description of the trapped environment is still incomplete, with largest uncertainties affecting the high energy fluxes in the inner zone and the South Atlantic Anomaly (SAA), where the inner belt makes its closest approach to the Earth.

In addition, the ma\-gne\-to\-sphe\-ric radiation includes populations of albedo protons originated by the collisions of interplanetary CRs on the atmosphere. A quasi-trapped component concentrates in equatorial regions and presents features similar to those of radiation belt protons, but with limited lifetimes and much less intense fluxes. An un-trapped component spreads over all la\-ti\-tu\-des including the ``penumbra'' region near the geo\-ma\-gne\-tic cutoff, where particles of both interplanetary and atmospheric origin are present.

In this Section we discuss the new accurate measurements of the ma\-gne\-to\-sphe\-ric radiation made by the PAMELA experiment \citep{PAMTRAPPED,ALBEDO}. Results are based on data collected between July 2006 and September 2009.

\subsection{Particle Classification}\label{Particle classification}
Trajectories of all selected down-going protons were reconstructed in the Earth's ma\-gne\-to\-sphe\-re using a tracing program based on numerical integration methods \citep{TJPROG,SMART}, and implementing the IGRF11 \citep{IGRF11} and the TS05 \citep{TS05} as internal and external geo\-ma\-gne\-tic field models, respectively \citep{BRUNO_ICRC_UNDERCUTOFF}. Solar wind and Interplanetay Magnetic Field (IMF) parameters were obtained from the high re\-so\-lu\-tion (5-min) Omniweb database (\url{http://omniweb.gsfc.nasa.gov/}). Trajectories were propagated back and forth from the measurement location and traced until: back-traced trajectories reached the model ma\-gne\-to\-sphe\-re boundaries (\emph{galactic} protons); or they intersected the absorbing atmosphere limit, which was assumed at an altitude\footnote{Such a value roughly corresponds to the mean production altitude for albedo protons.} of 40 km (\emph{re-entrant albedo} protons); or particles performed more than $10^{6}/R^{2}$ steps\footnote{Since the program uses a dynamic variable step length, which is of the order of 1\% of a particle gyro-distance in the magnetic field, such a criterion ensures that at least 4 drift cycles around the Earth were performed (2 cycles for each propagation direction).}, where $R$ is the particle rigidity (momentum/charge) in GV, for both propagation directions (\emph{geo\-ma\-gne\-tically trapped} protons). Trapped trajectories were verified to fulfil the adiabatic conditions, in particular the hierarchy of temporal scales: $\omega_{gyro} \gg \omega_{bounce} \gg \omega_{drift}$, where $\omega_{gyro}$, $\omega_{bounce}$ and $\omega_{drift}$ are the frequencies associated with gyration, bouncing and drift motions \citep{PAMTRAPPED}.

Albedo protons were classified into \emph{quasi-trapped} and \emph{un-trapped}. The former have trajectories similar to those of stably-trapped, but are originated and re-absorbed by the atmosphere during a time larger than a bounce period (up to several tens of s). The latter include both a short-lived component of protons \emph{precipitating} within a bounce period ($\lesssim$ 1s), and a long-lived (\emph{pseudo-trapped}) component with rigidities close to the geo\-ma\-gne\-tic cutoff (penumbra), cha\-ra\-cterized by a chaotic motion (non-adiabatic trajectories). Full details, including distributions of lifetimes and production/absorption points on the atmosphere, can be found in \citet{ALBEDO}.

\subsection{Flux Calculation}\label{Flux calculation}
Proton spectra were estimated by assuming an isotropic distribution in all the explored regions except the SAA. In fact, trapped fluxes are significantly anisotropic due to the interaction with the Earth's atmosphere, and thus the \emph{gathering power} of the apparatus depends on the spacecraft orientation with respect to the local geo\-ma\-gne\-tic field. In this case, a PAMELA \emph{effective area} (cm$^{2}$) was evaluated as a function of particle energy $E$, local pitch-angle $\alpha$ and satellite orientation $\Psi$:
\begin{equation}\label{area_formula}
H(E,\alpha,\Psi)=\frac{  sin\alpha}{2\pi}\int_{0}^{2\pi} d\beta \left[ A(E,\theta,\phi)  cos\theta \right],
\end{equation}
where $\beta$ is the gyro-phase angle, $\theta$=$\theta(\alpha,\beta,\Psi)$ and $\phi$=$\phi(\alpha,\beta,\Psi)$ are respectively the zenith and azimuth angles describing particle direction in the PAMELA frame\footnote{The PAMELA frame has the origin in the center of the spectrometer cavity; the Z axis is directed along the main axis of the apparatus, toward incoming particles; the Y axis is directed opposite to the main direction of the magnetic field inside the spectrometer; the X axis completes a right-handed system.}, and $A(E,\theta,\phi)$ is the apparatus response function. The effective area was evaluated with accurate simulations based on \citet{Sullivan}. Then, at given geographic location, differential directional fluxes were obtained as:
\begin{equation}\label{flux_formula}
F(E,\alpha)=\frac{N(E,\alpha)}{2\pi   \sum_{\Psi}  \left[H(E,\alpha,\Psi) \Delta T(\Psi)\right]   \Delta\alpha  \Delta E},
\end{equation}
where $N$ is the number of counts corrected for selection efficiencies, and the effective area is averaged over spacecraft orientations corresponding to the considered position, by weighting each $H(E,\alpha,\Psi)$ contribution by the livetime $\Delta T(\Psi)$ spent by PAMELA at orientation $\Psi$. At a later stage, the geographic flux grid was interpolated onto magnetic coordinates, including the adiabatic invariants and related variables. In particular, distributions were evaluated as a function of kinetic energy $E$, equatorial pitch-angle $\alpha_{eq}$ and McIlwain's $L$-shell, providing a convenient description of trapped fluxes. Further details can be found in \citet{PAMTRAPPED}.

\begin{figure}[!t]
\centering
\includegraphics[width=4.5in]{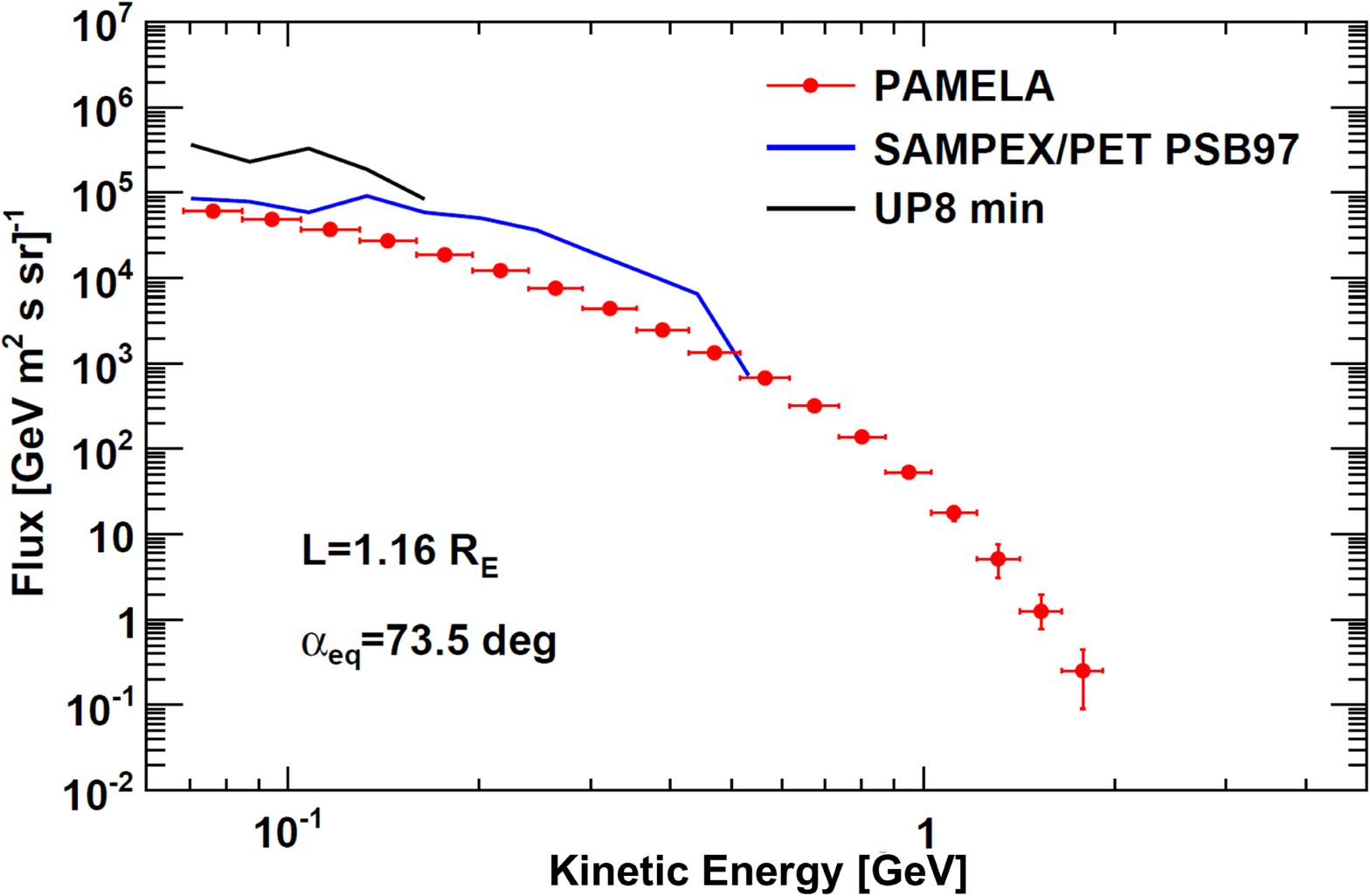}
\caption{PAMELA trapped proton energy spectrum for sample $\alpha_{eq}$ and $L$-shell values, compared with the predictions
from the UP8-min \citep{AP8} and the PSB97 \citep{PSB97} models (from SPENVIS, \url{https://www.spenvis.oma.be/}).}
\label{Fig2}
\end{figure}

\begin{figure}[!t]
\centering
\includegraphics[width=5.5in]{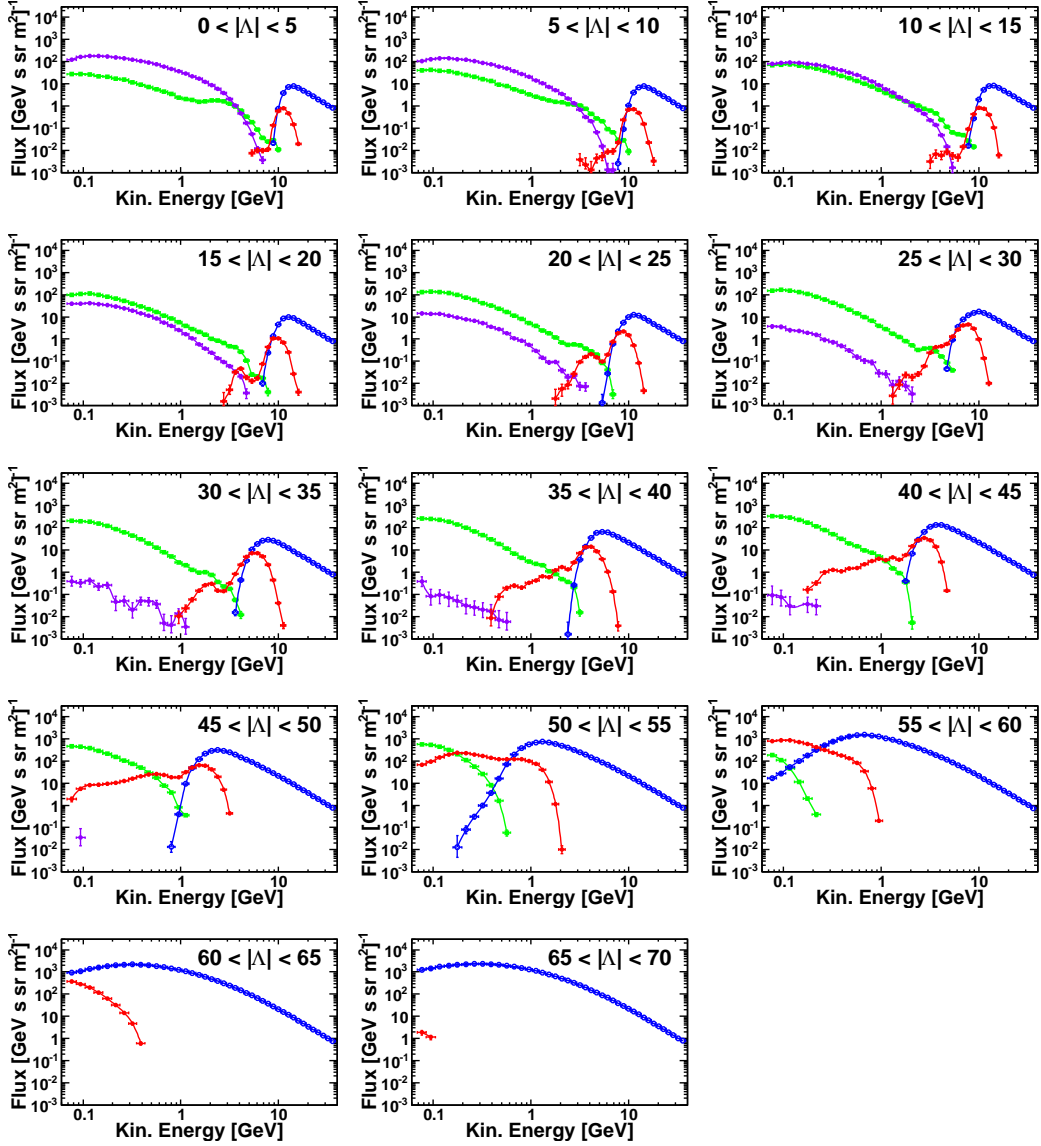}
\caption{Differential energy spectra outside the SAA for different bins of AACGM la\-ti\-tude $|\Lambda|$. Results for the several proton populations are shown: quasi-trapped (violet), precipitating (green), pseudo-trapped (red) and galactic (blue).}
\label{Fig3}
\end{figure}

\begin{figure}[!t]
\centering
\includegraphics[width=4.5in]{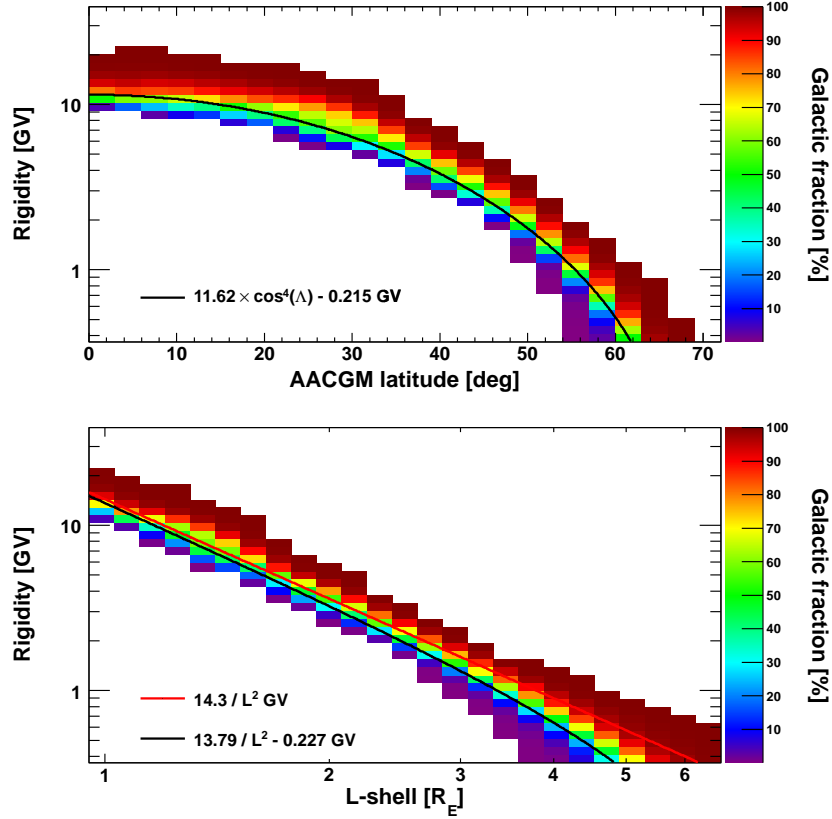}
\caption{Fraction of galactic protons in the penumbra region, as a function of particle rigidity and AACGM latitude $|\Lambda|$ (top) and $L$-shell (bottom). See the text for details.}
\label{Fig4}
\end{figure}

\subsection{Results}\label{Results}
PAMELA data extend the observational range for trapped protons down to $L$ $\sim$ 1.1 $R_{E}$, and up to the maximum kinetic energies cor\-res\-pon\-ding to the trapping limits (a few GeV). Figure \ref{Fig2} compares PAMELA geo\-ma\-gne\-tically trapped results (for sample $\alpha_{eq}$ and $L$ values) with the predictions from two empirical models available in the same energy and altitude ranges: the AP8 \citep{AP8} unidirectional (hereafter UP8) model for solar minimum conditions, and the SAMPEX/PET PSB97 model \citep{PSB97}. Model data were derived from the SPENVIS system (http://www.spenvis.oma.be/). In general, the UP8 model significantly overestimates PAMELA observations, while a better agreement can be observed with the PSB97 model. However, PAMELA fluxes do not show the spectral structures present in the PSB97 predictions.

Albedo fluxes were mapped using the Altitude Adjusted Corrected Geomagnetic (AACGM) coordinates \citep{HERES}, developed to provide a realistic description of high latitude regions accounting for the multipolar geo\-ma\-gne\-tic field. Figure \ref{Fig3} shows the spectra of the various albedo components outside the SAA (B$>$0.23 G) measured at different latitudes $|\Lambda|$, along with the galactic component. Fluxes were averaged over longitudes. Quasi-trapped protons are limited to low latitudes and to energies below $\sim$ 8 GeV; their fluxes smoothly decrease with increasing latitude and energy. Conversely, the precipitating component spreads to higher latitudes, with spectra extending up to $\sim$10 GeV. Finally, pseudo-trapped protons concentrate at highest latitudes and energies (up to $\sim$ 20 GeV), with a peak in the penumbra related to large gyro-radius ($10^{2}-10^{3}$ km) effects.

Features of the penumbra region are investigated in Figure \ref{Fig4}, where the fraction (\%) of galactic over total (galactic + albedo) pro\-tons is displayed as a function of particle rigidity, and AACGM latitude (top panel) and $L$-shell (bottom panel). The penumbra was identified as the region where both albedo and galactic proton trajectories were reconstructed. The black curves denote a fit of points with an equal percentage of the two components, while the red line refers to the St\"{o}rmer vertical cutoff for the PAMELA epoch.

%
%

\section{Solar Energetic Particles}
The PAMELA space experiment is providing first direct observations of SEPs in a large energetic interval ($\gtrsim$80 MeV) bridging the low energy measurements by in-situ spacecrafts and the ground level enhancement data by the worldwide network of neutron monitors. Its unique observational capabilities include the possibility of measuring the flux angular distribution and thus investigating possible anisotropies associated to SEP events. Results are supported by an accurate back-tracing analysis based on a realistic description of the Earth's ma\-gne\-to\-sphe\-re, which is exploited to estimate the SEP fluxes as a function of the asymptotic direction of arrival. In this Section we discuss the results for the May 17, 2012 event \citep{MAY17PAPER}.

\subsection{Asymptotic Directions of Arrival}
The asymptotic arrival directions (i.e. the directions of approach before encountering the ma\-gne\-to\-sphe\-re) of all selected interplanetary protons were e\-va\-lu\-a\-ted with the trajectory tracing method, based on the IGRF11 and the TS07D \citep{TS07D} models for the description of the internal and external\footnote{The TS07D model provides a larger spatial coverage with respect to the TS05 model.} geo\-ma\-gne\-tic field sources, respectively. The asymptotic directions were eva\-luated with respect to the IMF direction (based on Omniweb 5-min data), with polar angles $\alpha$ and $\beta$ denoting the pitch-angle and the gyro-phase angle.

\subsection{Flux Calculation}
Fluxes were reconstructed as a function of particle rigidity and asymptotic pitch-angle. The used approach is analogous to the one developed for the measurement of geo\-ma\-gne\-tically trapped protons (see Section \ref{Flux calculation}), but in this case the transformation between local ($\theta$,$\phi$) and magnetic ($\alpha$,$\beta$) angles depends on particle propagation in the magnetosphere, so the effective area was obtained through the trajectory tracing method. To assure a high re\-so\-lu\-tion, $\sim$2800 tra\-je\-cto\-ries (uniformly distributed inside PAMELA field of view) were reconstructed for 1-sec time steps along the satellite orbit and 22 ri\-gi\-di\-ty values between 0.39 -- 4.09 GV, for a total of $\sim8\times10^{7}$ trajectories for each polar pass. At a later stage, results were interpolated over the full field of view. Since PAMELA's semi-aperture is $\sim$20 deg, the observable pitch-angle range is relatively small (a few deg) except in the penumbral regions, where trajectories become chaotic  and cor\-res\-pon\-ding asymptotic directions rapidly change with particle rigidity and looking direction: due to the related measurement uncertainties, these zones we\-re excluded from the ana\-ly\-sis. Further details can be found in \citet{BRUNO_ICRC_SEP,BRUNO_ICPPA}.

\subsection{Results}
\begin{figure}[!t]
\centering
\includegraphics[width=5.5in]{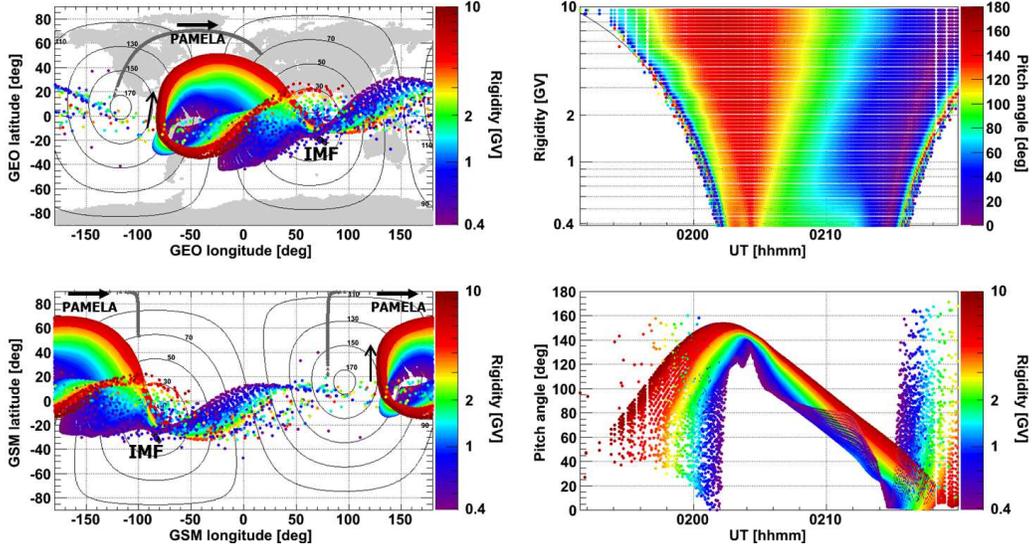}
\caption{PAMELA's vertical asymptotic directions of view (0.39 -- 10 GV) during the first polar pass that registered the May 17, 2012 SEP event. See the text for details.}
\label{Figure1}
\end{figure}

Figure \ref{Figure1} reports PAMELA's vertical asymptotic directions of view during the first polar pass that registered the May 17, 2012 SEP event. Left panels show the results in terms of GEO (top) and GSM (bottom) coordinates, for different rigidities (color codes). The spacecraft position is indicated by the dark grey curve. The contour curves represent values of constant pitch-angle with respect to the IMF direction, denoted with crosses. It can be noted that IMF was almost perpendicular to the sunward direction. As PAMELA was moving eastward and changing its orientation along the orbit, viewing asymptotic directions rapidly varied performing a clockwise loop (see the arrows). Right panels display the asymptotic directions as a function of UT, and particle rigidity (top) and pitch-angle (bottom); the color codes refer to the corresponding pitch-angle and rigidity values, respectively. Solid curves in the top-right panel denote the estimated St\"{o}rmer vertical cutoff. A large pitch-angle interval was covered during the polar pass (0--145 deg). In particular, PAMELA was looking at the IMF direction around 0216 UT.

\begin{figure}[!t]
\centering
\includegraphics[width=4.5in]{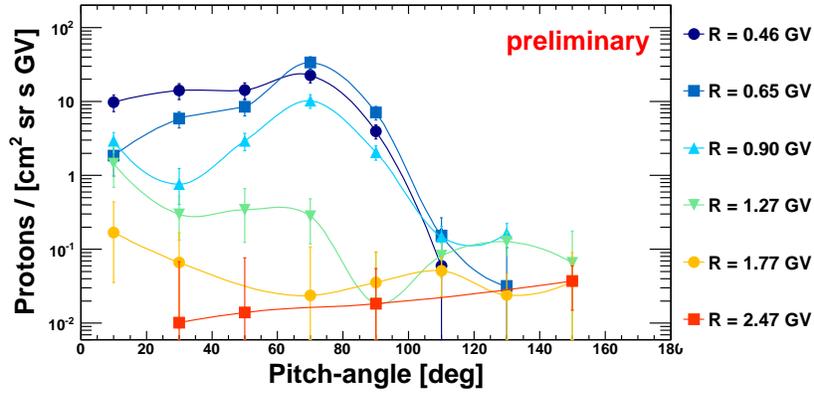}
\caption{SEP pitch-angle profiles measured by PAMELA (May 17, 0158 -- 0220 UT) for different rigidity bins (color code). Lines are to guide the eye.}
\label{Figure3}
\end{figure}

Preliminary pitch-angle profiles measured by PAMELA during the first polar pass are shown in Figure \ref{Figure3}, for different rigidity bins (color code). The galactic CR background, evaluated by using the data acquired by PAMELA during two days prior the SEP arrival, was subtracted from registered fluxes. The vertical error bars include only statistical uncertainties. Two populations with very different pitch-angle distributions can be noted: a low-energy component ($\lesssim$ 1 GV) confined to pitch-angles $<$90 deg and exhibiting si\-gni\-fi\-cant scattering or redistribution; and a high-energy component (1 -- 2 GV) that is beamed with pitch-angles $<$30 deg and relatively unaffected by dispersive transport effects, consistent with neutron monitor observations. The presence of these simultaneous populations can be explained by postulating a local scattering/redistribution in the Earth's magnetosheath (not included in the empirical models), with a major role played by the quasi-perpendicular IMF orientation (see \citet{MAY17PAPER} for a comprehensive discussion).

%
%

\section{Geomagnetic Cutoff Variations During Magnetospheric Storms}
The CR access to a specific location in the Earth's ma\-gne\-to\-sphe\-re is determined by the spatial structure and intensity of the geo\-ma\-gne\-tic field, which is an highly dynamical system: its configuration is driven by the solar wind and by the interaction between terrestrial and interplanetary fields, being compressed at the dayside and stretched toward the magnetotail on the nightside. Major space weather phenomena are caused by large SEP events. In case of earthward-directed Coronal Mass Ejections (CMEs) or co-rotating interaction regions disturbances can culminate in geo\-ma\-gne\-tic storms, characterized by a large transfer of the solar wind energy into the Earth's ma\-gne\-to\-sphe\-re, with significant changes in the currents, plasmas and fields.

In this Section we discuss the measurement of the cutoff variability during the strong geo\-ma\-gne\-tic storm on 14 December 2006, the last large CME-driven storm of solar cycle 23 \citep{BRUNO_ICRC_CUTOFF,GSTORM}.

\subsection{The 14 December 2006 Geomagnetic Storm}\label{The 14 December 2006 Geomagnetic Storm}
On 13 December 2006 at 0214 UT, an X3.4/4B solar flare occurred in the active region NOAA 10930 (S06W23). This event also produced a full-halo CME with a speed of 1774 km s$^{-1}$. The CME forward shock reached the Earth at $\sim$ 1410 UT on 14 December causing a Forbush decrease of galactic CR intensities that lasted for several days. In addition, the large increase in the solar wind velocity $V_{SW}$ caused a Storm Sudden Commencement (SSC). The increased dynamical pressure $P_{SW}$ resulted in a dramatic ma\-gne\-to\-sphe\-ric compression along with an intensification of the magnetopause current. The SSC marked the beginning of the initial phase of the storm, which was characterized by intense fluctuations in $P_{SW}$ and in all IMF components. In particular, $B_{z}^{IMF}$ became positive after 1800 UT on 14 December and it continued to oscillate until the $\sim$2300 UT, when the IMF intensity increased and $B_{z}^{IMF}$ rapidly turned negative, while $V_{SW}$ decreased. The main phase of the storm reached a maximum in the first hours of 15 December, followed by a slow ($\sim$ 3 days) recovery phase. The protracted large-amplitude southward IMF was associated with the magnetic cloud which caused the storm. Such large events are untypical of the intervals of low solar activity. An additional interplanetary shock, related to a less geo-effective CME, was registered on 16 December at $\sim$ 1800 UT.

\subsection{Evaluation of Geomagnetic Cutoff Latitudes}
The lowest magnetic latitude to which a charged CR particle can penetrate the Earth's ma\-gne\-tic field is known as its \emph{cutoff latitude} and is a function of particle rigidity, arrival direction and geo\-ma\-gne\-tic activity. Due to the narrow field of view of PAMELA, with its major axis mostly oriented toward the zenith, the measured fluxes correspond to approximately vertical directions.

The algorithm used to evaluate cutoff latitudes from the PAMELA data \citep{BRUNO_ICRC_CUTOFF} is si\-mi\-lar to one developed by \citet{LESKE2001}, using the low-energy proton and alpha particle measurements made by the SAMPEX mission. For each rigidity bin, a mean flux was obtained by averaging fluxes measured at latitudes higher than $\Lambda_{min}=cos^{-1} (R[GV]/20)^{1/4}$ deg, and the cutoff latitude was evaluated as the latitude where the flux intensity is equal to the half of the average value. $\Lambda_{min}$ represents an upper cutoff latitude derived from the experimental distributions to avoid penumbral effects. To support the analysis results, cutoff latitudes were also numerically modeled with back-tracing techniques: at a given rigidity, the \emph{modeled} cutoff latitude was evaluated as the latitude where reconstructed interplanetary and albedo flux intensities were equal.

The calculation was performed for 13 rigidity logarithmic bins in the range 0.39 -- 3.29 GV and the final cutoff values were derived by fitting PAMELA's data averaged over single orbital periods ($\sim$94 min), including two measurements (entering and exiting the polar caps) in both magnetic hemispheres.

\begin{figure*}[!t]
\centering
\includegraphics[width=4.5in]{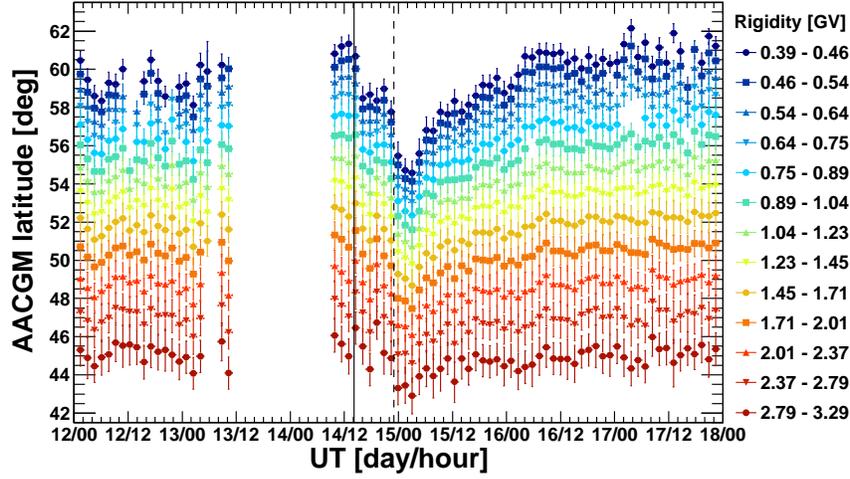}
\caption{Time profile of the geo\-ma\-gne\-tic cutoff latitudes measured by PAMELA, for different rigidity bins. Vertical solid and dashed lines mark the shock and the magnetic cloud arrival, respectively.}
\label{comparison}
\end{figure*}

\begin{figure*}[!t]
\centering
\includegraphics[width=4in]{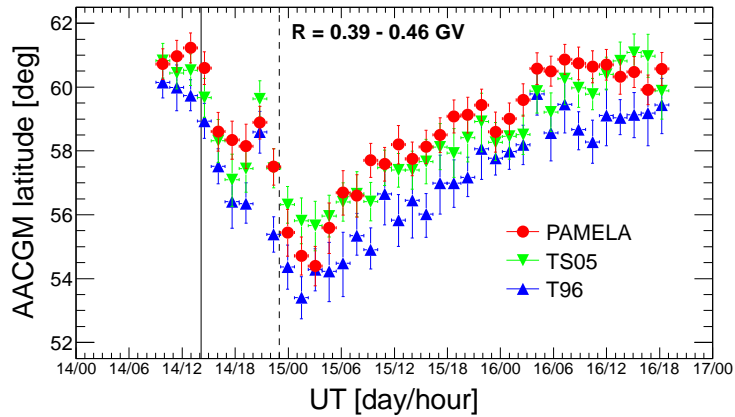}
\caption{Comparison between measured (red) and modeled (blue - T96 model; green - TS05 model) cutoff variations in the lowest rigidity interval. Vertical solid and dashed lines mark the shock and the magnetic cloud arrival, respectively.}
\label{modcomparison}
\end{figure*}

\subsection{Results}
Figure \ref{comparison} shows the geo\-ma\-gne\-tic cutoff AACGM latitudes measured by PAMELA as a function of time (12 -- 18 December 2006) for different rigidity bins (color code). Each point denotes the cutoff latitude value averaged over a single spacecraft orbit; the error bars include the statistical uncertainties of the measurement. Data were missed from 1000 UT on December 13 until 0914 UT on December 14 because of an onboard system reset of the satellite. The evolution of the December 14 magnetic storm followed the typical scenario in which the cutoff latitudes moves equatorward as a consequence of a CME impact on the ma\-gne\-to\-sphe\-re with an associated transition to southward $B_{z}^{IMF}$. The re\-gi\-stered cutoff variation decreases with increasing rigidity, with a $\sim$7 deg maximum suppression at lowest rigidities.

Figure \ref{modcomparison} reports the comparison between measured and modeled cutoff latitudes (0.39 -- 0.46 GV). Two different empirical descriptions of the external geo\-ma\-gne\-tic fields were employed (in combination with the IGRF11 model): the T96 \citep{T96} and the TS05 models. While the former appears to systematically underestimate (up to 4\%) the observations, PAMELA and TS05 results are in a good agreement within the statistical uncertainties. Further details, including a study of correlations between cutoff latitudes and main solar wind, IMF and geo\-ma\-gne\-tic (Kp, Dst, Sym-H) parameters can be found in \citet{GSTORM}.

\section{Conclusions}
This work reviews PAMELA's main ma\-gne\-to\-sphe\-ric results, with the focus on the ana\-ly\-sis methods developed to support the observations, based on an accurate reconstruction of particle trajectories in the Earth's ma\-gne\-to\-sphe\-re. The trajectory analysis enabled the separation of particle populations of atmospheric and interplanetary origin, improving the study of geo\-ma\-gne\-tically trapped particles from the inner radiation belt and of albedo protons at different latitudes, including the geo\-ma\-gne\-tic cutoff va\-ria\-tions during ma\-gne\-to\-sphe\-ric storms. The back-tracing approach was also exploited in the SEP analysis allowing the investigation of flux anisotropies, proving to be an important in\-gre\-dient for the interpretation of the solar events observed by PAMELA between solar cycles 23 and 24.

\section*{Acknowledgements}
\small
We acknowledge support from The Italian Space Agency (ASI), Deutsches Zentrum f$\ddot{u}$r Luftund Raumfahrt (DLR), The Swedish National Space Board, The Swedish Research Council, The Russian Space Agency (Roscosmos) and The Russian Scientific Foundation.



\end{document}